\documentclass[preprintnumbers,article,amsmath,amssymb,floatfix,10pt,prd,onecolumn,superscriptaddress,nofootinbib]{revtex4-2}

\usepackage{titlesec}
\usepackage[toc]{appendix}
\usepackage{caption}
\titleformat*{\section}{\LARGE\bfseries}
\titleformat*{\subsection}{\Large\bfseries}
\titleformat*{\subsubsection}{\large\bfseries}
\titleformat*{\paragraph}{\large\bfseries}
\titleformat*{\subparagraph}{\large\bfseries}
\usepackage{bm}
\usepackage{amsfonts}
\usepackage{latexsym}
\usepackage[latin1]{inputenc}
\usepackage{graphicx}
\usepackage{array} % for column alignment
\usepackage{adjustbox} % for fitting table to page width
\usepackage{amsmath}
\usepackage{palatino}
\usepackage{subcaption}
\usepackage{mathpazo}
\usepackage{textcomp}
\usepackage{booktabs}
\usepackage{dcolumn}
\usepackage{ragged2e}
\usepackage{hyperref}
\hypersetup{colorlinks,citecolor=blue}
\hypersetup{colorlinks=true,linkcolor=blue,filecolor=magenta,    urlcolor=blue}
\usepackage{amsthm}
\usepackage{xcolor}
\usepackage{orcidlink}
\usepackage{epsfig}
\usepackage{subcaption}
\usepackage{commath}
\usepackage{cancel, ulem}\usepackage{setspace}

\usepackage{multirow}
\newcommand{\m}{\mathring}

\def\jnl@style{\it}
\def\aaref@jnl#1{{\jnl@style#1}}

\def\aaref@jnl#1{{\jnl@style#1}}

\def\aj{\aaref@jnl{AJ}}                   % Astronomical Journal
\def\apj{\aaref@jnl{ApJ}}                 % Astrophysical Journal
\def\apjl{\aaref@jnl{ApJ}}                % Astrophysical Journal, Letters
\def\apjs{\aaref@jnl{ApJS}}               % Astrophysical Journal, Supplement
\def\apss{\aaref@jnl{Ap\&SS}}             % Astrophysics and Space Science
\def\aap{\aaref@jnl{A\&A}}                % Astronomy and Astrophysics
\def\aapr{\aaref@jnl{A\&A~Rev.}}          % Astronomy and Astrophysics Reviews
\def\aaps{\aaref@jnl{A\&AS}}              % Astronomy and Astrophysics, Supplement
\def\mnras{\aaref@jnl{Mon.~Not.~Roy.~Astron.~Soc.}}             % Monthly Notices of the RAS
\def\prd{\aaref@jnl{Phys.~Rev.~D}}        % Physical Review D
\def\prc{\aaref@jnl{Phys.~Rev.~C}}  % Physical Review C
\def\prl{\aaref@jnl{Phys.~Rev.~Lett.}}    % Physical Review Letters
\def\qjras{\aaref@jnl{QJRAS}}             % Quarterly Journal of the RAS
\def\skytel{\aaref@jnl{S\&T}}             % Sky and Telescope
\def\ssr{\aaref@jnl{Space~Sci.~Rev.}}     % Space Science Reviews
\def\zap{\aaref@jnl{ZAp}}                 % Zeitschrift fuer Astrophysik
\def\nat{\aaref@jnl{Nature}}              % Nature
\def\aplett{\aaref@jnl{Astrophys.~Lett.}} % Astrophysics Letters
\def\apspr{\aaref@jnl{Astrophys.~Space~Phys.~Res.}} % Astrophysics Space Physics Research
\def\physrep{\aaref@jnl{Phys.~Rep.}}      % Physics Reports
\def\physscr{\aaref@jnl{Phys.~Scr}}       % Physica Scripta
\def\commat{\aaref@jnl{Comm.~Math.~Phys.}}              % Communications in Mathematical Physics
\def\science{\aaref@jnl{Science}}               % Science
\def\cqg{\aaref@jnl{Classical Quant.~Grav.}}            % Classical and Quantum Gravity
\def\jpcs{\aaref@jnl{JPCS}}                                     % Journal of Physics Conference Series
\def\ijmpd{\aaref@jnl{Int.~J.~Mod.~Phys.~D}}                    % International Journal of Modern Physics D
\def\grg{\aaref@jnl{Gen.~Relat.~Gravit.}}               % General Relativity and Gravitation
\def\rpp{\aaref@jnl{Rep.~Prog.~Phys.}}          % Reports on Progress in Physics
\def\npa{\aaref@jnl{Nucl.~Phys.~A}}        % Nuclear Physics A
\def\lrr{\aaref@jnl{Living Rev.~Rel.}}                   % Living reviews in relativity
\def\jcap{\aaref@jnl{J.~Cosmology Astropart.~Phys.}}    % Journal of cosmology and astroparticle physics
\def\rmp{\aaref@jnl{Rev.~Mod.~Phys.}}   %Reviews of modern physics
\def\epjc{\aaref@jnl{Eur.~Phys.~J.~C}} 
\def\plb{\aaref@jnl{~Phy.~Lett.~B}} 
\def\mpla{\aaref@jnl{Mod.~Phy.~Lett.~A}} 
\def\arxiv{\aaref@jnl{arxiv.org}}

\allowdisplaybreaks[1]
\newcommand{\mD}{\mathring{\mathcal{D}}}
\begin{document}
\title{Phase-space analysis of an anisotropic universe in $f(Q,C)$ gravity}
\author{Ghulam Murtaza\orcidlink{0009-0002-6086-7346}}
\email{ghulammurtaza@1utar.my}
\affiliation{Department of Mathematical and Actuarial Sciences, Universiti Tunku Abdul Rahman, Jalan Sungai Long,
43000 Cheras, Malaysia}
\author{Avik De\orcidlink{0000-0001-6475-3085}}
\email{avikde@um.edu.my}
\affiliation{Institute of Mathematical Sciences, Faculty of Science, Universiti Malaya, 50603 Kuala Lumpur, Malaysia}
\author{Tee-How Loo\orcidlink{0000-0003-4099-9843}}
\email{looth@um.edu.my}
\affiliation{Institute of Mathematical Sciences, Faculty of Science, Universiti Malaya, 50603 Kuala Lumpur, Malaysia}
\author{Yong Kheng Goh\orcidlink{0000-0002-7338-9614}}\email{gohyk@utar.edu.my}
\address{Department of Mathematical and Actuarial Sciences,
Universiti Tunku Abdul Rahman, Jalan Sungai Long, 43000 Cheras, Malaysia}
\author {How Hui Liew\orcidlink{}}
\email{liewhh@utar.edu.my}
\affiliation{Department of Mathematical and Actuarial Sciences,
Universiti Tunku Abdul Rahman, Jalan Sungai Long, 43000 Cheras, Malaysia}
%\date{}
%\footnotetext{The research was supported by the Ministry of Higher Education (MoHE), through the Fundamental Research Grant Scheme (FRGS/1/2023/STG07/UM/02/3, project no.: FP074-2023).}
\footnotetext{The research has been carried out under Universiti Tunku Abdul Rahman Research Fund project IPSR/RMC/UTARRF/2023-C1/A09 provided by Universiti Tunku Abdul Rahman.}
\begin{abstract}
\textbf{Abstract}: In this study, we analyze the anisotropic universe in $f(Q,C)$ gravity theory. To achieve this, we consider three specific models of $f(Q,C)$ gravity and rewrite the equations of motion of each model as an autonomous system. We identify and analyze the critical points, examine their stability, and plot phase portraits to illustrate the behavior of each critical point. The evolution of key parameters, including the equation of state (EoS) parameter $w_{eff}$, the deceleration parameter $q$, and the standard density parameters $\Omega_{m}$ and $\Omega_{DE}$, is thoroughly investigated. The anisotropic dynamical variable exhibits decelerated behavior, consistent with the early universe, while the others demonstrate accelerated behavior, aligned with late-time observations across all models.
\end{abstract}

\maketitle
%\tableofcontents
\section{Introduction} \label{secc1}
At a sufficiently large scale, the universe is assumed to be homogeneous and isotropic, making it a spatially maximally symmetric FLRW geometry. This model is very convenient to use. However, this is an assumption on which the standard model of cosmology is built. However, the formation of stars, galaxies, and superclusters suggests that the universe was not isotropic, particularly in its early phases. There is considerable observational evidence suggesting that the universe exhibited asymmetries near the initial singularity \cite{Amirhashchi2017, Amirhashchi2018, Amirhashchi2019}.  Cosmic microwave background (CMB) \cite{akrami2020} indicates the existence of anisotropy and inhomogeneity in our universe especially in the early time of the universe also the cosmological observations of different phases of the universe (early and late time) reveal deviations in various physical parameters, raising concerns about the homogeneity and isotropy of the universe \cite{valentino2021}. The Wilkinson Microwave Anisotropy Probe (WMAP) data set strongly suggests that the standard FLRW model alone may not fully describe the structure of the universe. Since the standard model explains structure formation only when perturbed, this highlights the importance of reconsidering isotropy and incorporating an anisotropic spacetime. A notable finding in theoretical cosmology suggests that certain Bianchi models i.e I, VII0,V,VIIh, and XI can be used as the homogeneous limit of linear cosmological perturbations of the FLRW spacetime \cite{Pereira2019, Erickson2004}. By bridging the gap between the FLRW and Bianchi models, this finding encourages us to consider the latter as a viable interpretation of the standard cosmological model. These models reduce to  Friedmann spacetimes, when anisotropy approaches zero. Although the inflationary paradigm effectively transformed the early universe into its current homogeneous and isotropic state but still it cannot fully explain the entire evolutionary process. Therefore, to completely understand the history of cosmic evolution, one must go beyond the FLRW geometry and explore how the transition from an initially anisotropic and inhomogeneous state to the observed homogeneity and isotropy occurs. In this study, we consider the anisotropic yet homogeneous Bianchi type I metric, which is given by the line element \cite{Akarsu20199},
\begin{equation}\label{metric}
ds^2=-dt^2+a_1^2(t)dx^2+a_2^2(t)dy^2+a_3^2(t)dz^2.
\end{equation}
Here $a_i(t)$ represents the scale factor along $x$, $y$ and $z$ principle axes. This particular formulation provides the expansion factor in three different directions of the universe, which are orthogonal. By considering an anisotropic yet homogeneous background, the Bianchi Type I model introduces only minimal deviations from perfect isotropy and aligns more closely with modern large-scale structure and CMB observations, which place weak constraints on anisotropies \cite{Russell20144}. This allows Bianchi-I metric to be a good choice to study the various cosmological models that divert from perfect isotropy, as extensively studied \cite{adhav2012,sharif2011,farasat2016,campanelli2007,sarmah2023,jaffe2005,jaffe2006,jaffe2006a,sarmah2022,esposito2022,amirhashchi2020,bolejko2011,barrow1997,szekeres1975}.\\\\
 The general theory of relativity (GR) has been verified by the number of observations from millimeter scale to the solar system tests. CMB \cite{akrami2020} and la supernovae \cite{riess1998,perlmutter1999} observations support the faster expansion of the universe than as it used to be. So by the advancement in observational cosmology, it is widely recognized that GR is not able to explain the late-time accelerated expansion of the universe without considering the cosmological constant in its dynamics.  The searches regarding the existence of dark energy thus far have not had any worthwhile success. Additionally, in the context of GR some theoretical \cite{Helbig2020,rajantie2012,hawking1966} and observational \cite{buchert2016,socas2019} disagreement including some prominent examples like Hubble tension \cite{valentino2021,aghanim2018,chudaykin2021}, $\sigma_{8}$ tension \cite{aghanim2018,Poulin2023}, fine tune and coincident problem \cite{velten2014,zheng2021} are still unsettled. \\\\
 To address these issues, geometric modifications to GR, called modified theories of gravity, have been proposed. $f(\mathring R)$ theory is one of the simplest modified theory of GR in which the Ricci scalar $\mathring{R}$ computed from the Levi-Civita connection $\mathring{\Gamma}$ in Einstein-Hilbert action of GR is replaced with an arbitrary function $f(\mathring{R})$. Using this curvature-based approach to geometry, many modified theories were formulated \cite{saridakis2021,bohmer2021}.\\\\
 Recently, a class of modified gravity theory was developed by considering a more general affine connection over the standard use of the Levi-Civita connection \cite{bohmer2023,iosifidis2023} and abandoning the metric compatibility and/or torsion-free conditions. One of the most effective gravity theories of this class is the $f(Q)$ theory \cite{jimenez2018,jimenez2020,lymperis2022}. The scalar field $Q$ is the non-metricity scalar and it deviates from the Ricci Scalar $\mathring{R}$ by the boundary term $C$ as given in Section \ref{sec1}. Following the same mechanism as $f(\mathring{R})$, $f(Q)$ is a modified GR construction. The linear function $f(Q)=\alpha Q+\beta$ ultimately restores GR and it differs by boundary term at the level of Lagrangian \cite{Capozziello2022}. That is why this particular linear case is also called the symmetric teleparallel equivalent of GR (STEGR). Since $f(Q)$ theory is of second order as in GR, to increase the order of symmetric teleparallel theories, one can introduce higher order terms like $\Box Q, \Box^{k} Q $ in the Lagrangian \cite{otalora2016}. However, a more natural way is to include the boundary term $C$ in the Lagrangian to develop $f(Q,C)$ gravity as very recently proposed in \cite{avik2024, Capozziello2023a}. Several cosmological aspects of $f(Q,C)$ theory have been studied, in both theoretical and observational grounds, for reference see \cite{samaddar2025, Sadatian2024, Sharif2025, Alruwaili2025, Mushtaq2025, Bhoyar2025, Myrzakulov2025, muhammad2024, Chandra2024, Zotos2024, mmaurya2024, ppaliathanasis2024, Muhammad2024, maurya2024, De2023, Paliathanasiss2024, Dixit2024}. The scalar perturbation formulation and the growth structure of the $f(Q,C)$ theory have been studied \cite{Subramaniam2024}.\\\\
 In the context of metric-affine theories, symmetric teleparallel gravity and its extensions are becoming increasingly influential in the pursuit of a unified theory of gravity. Symmetric teleparallel theories and metric teleparallel and their extensions $f (T)$ \cite{Bahamonde2023a} and $f (Q)$ \cite{Heisenberg2024} theories are quite successful in explaining the latest observational aspects. Several cosmological studies have been done in the recent past \cite{lu2019,khyllep2021,mandal2020,hassan2021,barros2020,anagnostopoulos2021,atayde2021, Saha2025, Nashed2025, Loo2025,Terzis2025,Kavya2024,Wang2024,Gadbail2024,Ren2024,Gonalves2024}. However, most works so far have only considered the homogeneous and isotropic background geometry. Only a very limited number of works have considered the contribution of anisotropy in this theory \cite{sarmah2023,de2022,devi2022}. The cosmological studies of various anisotropic models in other curvature based modified theories can be found in \cite{Liu2017, Bishi2017, Shamir2010, Sharif2010, Reddy2012, Ahmed2014, Sahoo2015}.\\\\
 In cosmology, we generally deal with the cosmological field equations which are a set of coupled non-linear ordinary differential equations. Since there is no general way to solve non-linear differential equations. So dynamical system analysis (DSA) can be a good choice for doing a qualitative analysis of the evolution of such non-linear systems. To formulate the dynamical system, initially, a set of suitable dimensionless normalized dynamical variables are defined so that the field equations can be expressed as a closed set of coupled first-order non-linear differential equations. The trajectories in phase space help map out how the universe might transition from one era to another.  The fixed points in the phase space often correspond to important cosmological epochs like inflation, radiation/matter domination, or late-time acceleration.
Therefore, to understand cosmic evolution and its dynamics in cosmology and modified gravity models, DSA has proven to be an effective technique \cite{Bahamonde2018b, Wainwright1997,Coley2013}. By investigating the stability conditions, it allows one to theoretically constrain the range of viable models. A comprehensive DSA for $f(\mathring R)$ gravity was carried out in \cite{Odintsov2017a}. DSA for analyzing the interaction between dark matter and dark energy in the framework of $f(\mathring R)$ gravity has been explored in \cite{Roy2024, Chaterje2024}. A model-independent approach of DSA to study cosmology in  $f(\mathring R)$ gravity models has been discussed in \cite{Louw2024}. DSA of the isotropization of a pre-bounce contracting phase in $f(\mathring R)$ gravity was given in \cite{Aroraa2022}. In $f(\mathring R,\mathbb T)$ gravity, as an extension of $f(\mathring R)$ gravity by combining the trace of stress-energy tensor $\mathbb T$ was proposed and later DSA was performed to study the cosmological background evolution of the scalar-tensor representation of it. As discussed in \cite{Lobo2024}, aiming to identify dynamical cosmological behaviors that align with the $\Lambda$CDM model without requiring a dark energy component. DSA to explore the late-time cosmological evolution within a general class of $f(\mathring R,\mathbb T)$ gravity models featuring minimal curvature-matter coupling can be found in \cite{Ziaie2017}. A comprehensive DSA and some other notable works in $f(\mathring R,\mathbb T)$ gravity can be studied in \cite{Bhatti2022,mirzaa2016,farhoudi2013}. In metric teleparallel theory, DSA was conducted in two accelerating models of $f(T)$ theory \cite{duchaniya2023,duchaniya2022} and later to constrain $f(T)$ gravity models in \cite{Mirzaa2017}. DSA for scalar field potentials and certain cosmological models in $f(T)$ gravity can be found in \cite{Tripathy2024, Sarkar2025}. DSA of the background and perturbations in $f(Q)$ gravity was performed for both the exponential and the power-law models of $f(Q)$ in \cite{Khyllep2023a}. In both cases, the analysis revealed a matter-dominated saddle point with the correct growth rate of matter perturbations. This phase is followed by a smooth transition to a stable, dark energy-dominated accelerated universe, where matter perturbations remain constant. DSA of scalar field cosmology in coincident $f(Q)$ gravity was carried out in \cite{Ghosh2024b}. Phase space analysis of some $f(Q)$ theory models involving three different affine connections in an isotropic universe can be found in \cite{Shabani2023a}. DSA of two models of $f(Q)$, the power-law model $f(Q) = Q+mQ^{n}$ and logarithmic model $f(Q) = \alpha + \beta \log Q$ has been carried out in \cite{shah2023}. Both models lead to an accelerating and stable universe characterized by constant matter perturbations. Additionally, the power-law model results in a matter-dominated saddle point with the correct matter perturbation growth rate. In contrast, the logarithmic model leads to a saddle point dominated by the geometric component of the model, with perturbations in the matter sector. A detailed analysis of the power exponential model $f(Q) = Q e ^{\lambda \frac{Q_{0}}{Q}}$ using the DSA in $f(Q)$ theory has been carried out, considering two fluid components: radiation and matter, in \cite{Lazkoz}. For the model parameter $\lambda \not =0$, the existence of a radiation-dominated early time epoch, a saddle point dominated by matter, and an accelerating de Sitter attractor has been found. Several other recent and significant studies in $f(Q)$ theory using DSA approach are presented in \cite{Channuie2024,solanki2024,Mahataa2024,narawade2023,Paliatha2023,alam2023,Koussour2024}. DSA work in teleparallel theory with the boundary term, $f(T,B)$ was explored in \cite{franco2020, Kadam2023b}. Similar work in symmetric teleparallel theory with the boundary term, $f(Q,C)$ was examined in \cite{Lohakare2024, Shabani2024a}. However, all these works only considered the homogeneous and isotropic universe. Recently DSA of the LRS-BI universe with $f(Q)$ gravity theory was published \cite{Goswami2024}, however, severe additional constraints were imposed on it. In the context of metric teleparallel theories, DSA was attempted in the presence of anisotropy only in \cite{BI fTB}. In this article, we perform the dynamical system analysis to study the Bianchi-I cosmology in $f(Q,C)$ gravity.\\\\
This paper is organized in the following way: after the introductory Section \ref{secc1}, we discuss the basic formalism of $f (Q,C)$ gravity in Section \ref{sec1}, followed by the formulation of the Bianchi-I cosmology in $f (Q,C)$ theory in Section \ref{sec3}. Finally, we derive the general dynamical system for Bianchi-I cosmology under $f (Q,C)$ theory in Section \ref{sec4} and in its subsections, we perform dynamical system analysis on the three models of $f (Q,C)$ theory. We conclude our findings in Section \ref{sec5}.

%%%%%%%%%%%%%%%%%%%%%%%%%%%%%%%%%%%%%%%%%%%%%%%%%%%%%%%%%%%%%%%%%%%%%%%%%%%%
\section{Basic formalism of $f(Q,C)$ gravity}\label{sec1}
Let us start with the general formulation of $f(Q,C)$ gravity known as teleparallel geometries. In general, a metric affine manifold is composed of a four-dimensional Lorentzian manifold, represented as $M$, with its metric tensor $g_{\mu\nu}$ and the covariant derivative $\nabla_\lambda$ that is determined by an affine-connection $\Gamma^{\alpha}_{\,\,\,\mu\nu}$. By considering the two particular conditions on the connection, that is, metric compatibility and torsion-free, we only have one connection known as the Levi-Civita connection $\mathring{\Gamma}^\alpha{}_{\mu\nu}$ and it is commonly related to the  metric $g$ as, 
\begin{equation}
\mathring{\Gamma}^\alpha_{\,\,\,\mu\nu}=\frac{1}{2}g^{\alpha\beta}\left(\partial_\nu g_{\beta\mu}+\partial_\mu g_{\beta\nu}-\partial_\beta g_{\mu\nu}  \right)\,.
\end{equation}
Here the triplet $(M,g_{\mu\nu},\mathring{\Gamma}^\alpha_{\,\,\,\mu\nu})$ represents a Riemannian geometry. This Levi-Civita connection is a dependent contributor to the geometry of spacetime and also a function of $g_{\mu\nu}$. We get a significant development, by easing these conditions and adopting the torsionless and curvature less affine condition $\Gamma^\alpha{}_{\mu\nu}$ given as,
\begin{align}
\mathbb T^\alpha{}_{\mu\nu}:=&
%\Gamma^{\alpha}{}_{\nu\mu}-\Gamma^\alpha{}_{\mu\nu}=0\,, \quad 
2\Gamma^{\alpha}{}_{[\nu\mu]}=0\,,
\label{eqn:torsion-free}\\
R^\lambda{}_{\mu\alpha\nu}:=&
%\partial_\alpha\Gamma^\lambda{}_{\mu\nu}-\partial_\nu\Gamma^\lambda{}_{\mu\alpha}
%+\Gamma^\lambda{}_{\sigma\alpha}\Gamma^\sigma{}_{\mu\nu}
%-\Gamma^\lambda{}_{\sigma\nu}\Gamma^\sigma{}_{\mu\alpha}=0\,, 
2\partial_{[\alpha}\Gamma^\lambda{}_{|\mu|\nu]}
+2\Gamma^\lambda{}_{\sigma[\alpha}\Gamma^\sigma{}_{|\mu|\nu]}=0\,, 
\label{eqn:curvature-free}
\end{align}
leading us to symmetric teleparallel geometry. In this geometry, the term ``teleparallel'' refers to the parallel transport that is not dependent on the path, which is described by the covariant derivative and its related affine connection because of the vanishing Riemannian curvature tensor. Furthermore, the affine connection possesses the symmetric property in its lower indices due to the torsion-free and curvature-free conditions on it, and therefore the term ``symmetric'' is utilized. The non-metricity tensor that describes how this affine connection is not compatible with the metric is described as,
\begin{equation} \label{Q tensor}
Q_{\lambda\mu\nu} := \nabla_\lambda g_{\mu\nu}=\partial_\lambda g_{\mu\nu}-\Gamma^{\beta}_{\,\,\,\mu\lambda}g_{\beta\nu}-\Gamma^{\beta}_{\,\,\,\nu\lambda}g_{\beta\mu}\neq 0 \,.
\end{equation}
One can always write
\begin{equation} \label{connc}
\Gamma^\lambda{}_{\mu\nu} := \mathring{\Gamma}^\lambda{}_{\mu\nu}+L^\lambda{}_{\mu\nu}~,
\end{equation}
where $L^\lambda{}_{\mu\nu}$ represents the disformation tensor.
It can be given as,
\begin{equation} \label{L}
L^\lambda{}_{\mu\nu} = \frac{1}{2} (Q^\lambda{}_{\mu\nu} - Q_\mu{}^\lambda{}_\nu - Q_\nu{}^\lambda{}_\mu) \,.
\end{equation}
One can formulate two different kind of non-metricity vectors,
\begin{equation*}
 Q_\mu := g^{\nu\lambda}Q_{\mu\nu\lambda} = Q_\mu{}^\nu{}_\nu \,, \qquad \tilde{Q}_\mu := g^{\nu\lambda}Q_{\nu\mu\lambda} = Q_{\nu\mu}{}^\nu \,.
\end{equation*}
Similarly, one can write
\begin{align}
 L_\mu := L_\mu{}^\nu{}_\nu \,, \qquad 
 \tilde{L}_\mu := L_{\nu\mu}{}^\nu \,.   
\end{align}
We can express the superpotential or also known as non-metricity conjugate tensor $P^\lambda{}_{\mu\nu}$ as
\begin{equation} \label{P}
P^\lambda{}_{\mu\nu} = 
%\frac{1}{4} \left( -2 L^\lambda{}_{\mu\nu} + Q^\lambda g_{\mu\nu} - \tilde{Q}^\lambda g_{\mu\nu} -\frac{1}{2} \delta^\lambda_\mu Q_{\nu} - \frac{1}{2} \delta^\lambda_\nu Q_{\mu} \right) \,.
\frac{1}{4} \left( -2 L^\lambda{}_{\mu\nu} + Q^\lambda g_{\mu\nu} - \tilde{Q}^\lambda g_{\mu\nu} -\delta^\lambda{}_{(\mu} Q_{\nu)} \right) \,.
\end{equation}
So we can define the non-metricity scalar $Q$ in following way
\begin{equation} \label{Q}
Q=Q_{\alpha\beta\gamma}P^{\alpha\beta\gamma}\,.
\end{equation}
We can get more relations, following the two constraints (\ref{eqn:torsion-free})--(\ref{eqn:curvature-free}) which we have obtained during the derivation of curvature tensor associated with the Levi-Civita connection:
\begin{align}
\m R_{\mu\nu}+\m\nabla_\alpha L^\alpha{}_{\mu\nu}-\m\nabla_\nu\tilde L_\mu
+\tilde L_\alpha L^\alpha{}_{\mu\nu}-L_{\alpha\beta\nu}L^{\beta\alpha}{}_\mu=0\,,
\label{mRicci}\\
\m R+\m\nabla_\alpha(L^\alpha-\tilde L^\alpha)-Q=0\,. \label{mR}
\end{align}
As $Q^\alpha-\tilde Q^\alpha=L^\alpha-\tilde L^\alpha$, using the earlier relation, we can also express the boundary term as
\begin{align}
C=\m{R}-Q&=-\m\nabla_\alpha(Q^\alpha-\tilde Q^\alpha)%\notag\\
=-\frac1{\sqrt{-g}}\partial_\alpha\left[\sqrt{-g}(Q^\alpha-\tilde Q^\alpha)\right].
\end{align}
The action is given as
\begin{equation}
S=\int \left[ \frac{1}{2\kappa }f(Q,C)+\mathcal{L}_{M}\right] \sqrt{-g}%
\,d^{4}x\,,
\label{eqn:action-fQC}
\end{equation}
where $f$ is a function of $Q$ and $C$; and $\mathcal L_m$ denotes the matter Lagrangian.
The field equation we obtain by varying this action:
\begin{align}
\kappa T_{\mu\nu}
=&-\frac f2g_{\mu\nu}
  +\frac2{\sqrt{-g}}\partial_\lambda \left(\sqrt{-g}f_Q P^\lambda{}_{\mu\nu}
  \right)
  +(P_{\mu\alpha\beta}Q_\nu{}^{\alpha\beta}-2P_{\alpha\beta\nu}Q^{\alpha\beta}{}_\mu)
        f_Q \nonumber\\
&
  +\left(\frac C2 g_{\mu\nu}
  +\mD_{\mu\nu}
  -2P^\lambda{}_{\mu\nu}\partial_\lambda \right)f_C\,.
\label{eqn:FE1-pre}
\end{align}
We can re-write the field equation of the preceding connection by considering the coincident gauge as
\begin{align}\label{eqn:FE2b}
\partial_\mu\partial_\nu\left(\sqrt{-g}
\left[4(f_Q-f_C)P^{\mu\nu}{}_\lambda+\Delta_\lambda{}^{\mu\nu}\right]
\right)=0\,,
\end{align}
where
$$\mD_{\mu\nu}
:=-\m\nabla_{\mu}\m\nabla_{\nu}+g_{\mu\nu}\m\nabla^\alpha\m\nabla_\alpha\,,
$$
and 
$\Delta_\lambda{}^{\mu\nu}=-\frac2{\sqrt{-g}}
\frac{\delta(\sqrt{-g}
\mathcal L_m)}{\delta\Gamma^\lambda{}_{\mu\nu}}$ is hypermomentum tensor \cite{hyper}. In addition, the covariant form of the metric field equation (\ref{eqn:FE1-pre}) is given by
\begin{align}\label{FE2}
\kappa T_{\mu\nu}
&=-\frac f2g_{\mu\nu}+2P^\lambda{}_{\mu\nu}\nabla_\lambda(f_Q-f_C)
  +\left(\m G_{\mu\nu}+\frac Q2g_{\mu\nu}\right)f_Q
  +\left(\frac C2g_{\mu\nu}
  +\mD_{\mu\nu}
   \right)f_C\,.
\end{align}
The effective energy-momentum tensor can be defined as 
\begin{equation} \label{T^eff}
 T^{\text{eff}}_{\mu\nu} =  T_{\mu\nu}+ \frac 1{\kappa}\left[\frac f2g_{\mu\nu}-2P^\lambda{}_{\mu\nu}\nabla_\lambda(f_Q-f_C)
  -\frac {Qf_Q}2g_{\mu\nu}-\left(\frac C2g_{\mu\nu}
  +\mD_{\mu\nu}
  \right)f_C\right]\,,
\end{equation}
to get the equation as in GR
\begin{align}
    \m G_{\mu\nu}=\frac{\kappa}{f_Q}T^{\text{eff}}_{\mu\nu}\,.
\end{align}
We can see the additional term in (\ref{T^eff}), which appears due to the modification performed in the geometry to formulate the $f(Q,C)$ theory, serving as a dark energy like component:
\begin{align}
    T^{\text{DE}}_{\mu\nu}= \frac 1{f_Q}\left[\frac f2g_{\mu\nu}-2P^\lambda{}_{\mu\nu}\nabla_\lambda(f_Q-f_C)
  -\frac {Qf_Q}2g_{\mu\nu}-\left(\frac C2g_{\mu\nu}
  +\mD_{\mu\nu}
   \right)f_C\right]\,.
\end{align}
This $T^{\text{DE}}_{\mu\nu}$ part, as a key component of modified gravity theories and induces the negative pressure to acquire the late time acceleration.\\\\
An anisotropic spacetime metric is assumed to examine the non-trivial isotropization in the universe's evolution, and the equation of state (EoS) parameter of gravitational fluid is also generalized to get a more suitable anisotropic model. Since the universe is isotropized, the fluid also becomes isotropic, resulting in zero isotropic pressure and a vanishing skewness parameter. The energy-momentum tensor of anisotropic fluid is given as
\begin{equation} \label{T}
T^\mu_\nu = \text{diag}(-\rho, p_x, p_y, p_z) \,,
\end{equation}
where $\rho$ represents the energy density of the fluid, while  $p_x$, $p_y$ and $p_z$ denote the pressure in $x$, $y$ and $z$ directions, which are characterized by their respective EoS parameters  $\omega_1$, $\omega_2$ and $\omega_3$.
%%%%%%%%%%%%%%%%%%%%%%%%%%%%%%%%%%%%%%%%%%%%%%%%%%%%%%%%%%%%%%%%%%%%%%%%%%%%
\section{Bianchi-I cosmology in $f(Q,C)$ gravity}\label{sec3}
The anisotropic Bianchi-I metric in Cartesian coordinates in the present context is given by (\ref{metric}), we follow the framework of \cite{77,76} to set up the dynamical system to be analyzed in the following sections. The directional Hubble parameters are given by $H_i=\frac{\dot{a_i}}{a_i}$. We denote the arithmetic mean of these directional Hubble parameters by 
\begin{align}
    H(t)=\frac 13\left[H_1+H_2+H_3\right].
\end{align}
We also denote $a(t)$, by the average scale factor as the geometric mean $ a(t)=\left[a_1(t)a_2(t)a_3(t)\right]^{\frac13}$. By using the parametrization $a_i(t)=a(t)e^{\beta_i(t)}$, then we obtain $ H_i=H+\dot{\beta}_i, where~ \beta_1+\beta_2+\beta_3=0$. Obviously, $\dot{\beta}_i=0$ reduces the system to spatially homogeneous and isotropic FLRW. We define the anisotropy parameter $\sigma$ as
\begin{align}
    \sigma^2=\sum_{i=1}^3 \dot \beta_i^2.
\end{align}
We can compute the non-metricity scalar $Q$, the Ricci scalar $\mathring{R}$ and the boundary term $C$ as 
\begin{align}
Q&=-2\sum_{j<k}H_jH_k=-6H^2+\sigma^2,
\end{align}
\begin{align}
\mathring{R}= 2\sum_{i}(\dot{H}_i+H_i^2) + 2\sum_{j<k}H_jH_k =6\dot{H}+12H^{2}+\sigma^{2},
\end{align}
\begin{align}
C=\mathring{R}-Q= 6\dot{H}+18H^{2}.
\end{align}
We get the following equations of motion by using the field equations (\ref{FE2}):
\begin{align} 
\kappa \rho 
=& \frac f2 +2\sum_{j<k}H_jH_k f_Q-\left[\sum_i(\dot H_i+H_i^2)+2\sum_{j<k}H_jH_k\right]f_C
    +\sum_iH_i\dot f_C~,\\
\kappa p_1 
=&-\frac f2-\left[3H(H_2+H_3)+\dot{H}_2+\dot{H}_3\right]f_Q
    -\left[ H_2+H_3\right]\dot{f}_Q-\ddot{f}_C \notag\\
 & +\left[\dot{H}_1+\dot{H}_2+\dot{H}_3
    +2(H_1H_2+H_1H_3+H_2H_3)\right]f_C\,, \label{eom2}
\\  %\end{align}\begin{align}
\kappa p_2 
=&-\frac f2-\left[3H(H_1+H_3)+\dot{H}_1+\dot{H}_3\right]f_Q
    -\left[ H_1+H_3\right]\dot{f}_Q-\ddot{f}_C \notag\\
 & +\left[\dot{H}_1+\dot{H}_2+\dot{H}_3
    +2(H_1H_2+H_1H_3+H_2H_3)\right]f_C
\,,\label{eom3}
\\ %\end{align}\begin{align}
\kappa p_3 
=&-\frac f2-\left[3H(H_1+H_2)+\dot{H}_1+\dot{H}_2\right]f_Q
    -\left[ H_1+H_2\right]\dot{f}_Q-\ddot{f}_C \notag\\
 & +\left[\dot{H}_1+\dot{H}_2+\dot{H}_3
    +2(H_1H_2+H_1H_3+H_2H_3)\right]f_C
\,.\label{eom4}
\end{align}
The above set of equations of motion can be combined as the following:
\begin{align}
    \kappa \rho -\frac f2 
=&\left[6H^2-\sigma^2\right]f_Q-\left[3\dot H+9H^2\right]f_C+3H\dot f_C,\\
  \kappa p+\frac{f}{2}=&-\ddot{f_{C}}-2H\dot{f_{Q}}+(3\dot{H}+9H^{2})f_{C}-f_{Q}(2\dot{H}+6H^{2}),
\end{align}
\begin{align}
    \dot{\sigma}=-\sigma(\frac{\dot{f}_{Q}}{f_{Q}}+3H) .    
\end{align}
%%%%%%%%%%%%%%%%%%%%%%%%%%%%%%%%%%%%%%%%%%%%%%%%%%%%%%%%%%%%%%%%%%%%%%%%%%%%
\section{Dynamical system formulation for Bianchi-I universe in $f(Q,C)$ gravity}\label{sec4}
We consider the universe to be filled with matter fluid with energy density for the matter $\rho_{m}$. Since it is a matter-dominant universe so $p_{m}=0$ and $\omega_{m}$ vanishes. So, we  choose the following dynamical variables:
\[
X = f_{C}, \quad Y = f_{C} \frac{\dot{H}}{H^{2}}, \quad Z = f'_{C}, \quad 
% V = \frac{\kappa \rho_{r}}{3H^{2}}, 
W = -\frac{f}{6H^{2}}, \quad M = \frac{\sigma^{2}}{3H^{2}}.
\]
The usual expressions for matter and dark energy density parameters are,
\[
\Omega_{m} = \frac{\kappa \rho_{m}}{3H^{2}}, % \Omega_{r} = \frac{\kappa \rho_{r}}{3H^{2}}, 
\quad \Omega_{DE} = \frac{\kappa \rho_{DE}}{3H^{2}}
,\]
%\begin{align}
% \Omega_{m}+ \Omega_{r}+ \Omega_{DE}=1
%\end{align}
so the constraint equation is given as,
\begin{align}
 \Omega_{m} %\Omega_{r}
 -2f_{Q}+Mf_{Q}+W-Z+3X+Y=0.
\end{align}
%where
%\begin{align}
 %\Omega_{DE}=1-2f_{Q}+Mf_{Q}+W-Z+3X+Y
%\end{align}
To formulate the general dynamical system, we take the derivative of the dynamical variables with respect to cosmic time $t$ and then use the Hubble-normalized dimensionless time variable $N = \ln a$ to get the autonomous form:
\begin{align}
  X' &= Z, \label{e:ads1}\\
  Y' &=\lambda X+\frac{Y(Z-2Y)}{X}, \\
%\begin{align}
 %V'=-4V-\frac{2YV}{X}
%\end{align}
 W' &= -\frac{2WY}{X}-\lambda X+\frac{2Yf_{Q}}{X}-6Y+Mf'_{Q}+3Mf_{Q} ,\\
%\begin{align}
 %Z'=-2f'_{Q}-\frac{2Yf_{Q}}{X}-3Mf_{Q}+3Z+\frac{2Y}{X}-\frac{YZ}{X}
%\end{align}
 Z' &= -2f'_{Q}-\frac{2Yf_{Q}}{X}-6f_{Q}+3Y+9X+3W-\frac{YZ}{X}, \\
 M' &= -\frac{2Mf'_{Q}}{f_{Q}}-6M-\frac{2YM}{X} . \label{e:ads6}
\end{align}

Here, the prime represents the derivative with respect to $N$. Note that, we have used the parameter $\lambda=\frac{\ddot{H}}{H^3}$ \cite{franco2020,Odintsov2018} to form our autonomous dynamical system. In the latter reference, it was analyzed that for some particular constant value of $\lambda$, some cosmological solutions can be restored. For instance, for $\lambda=0$ we can obtain the de Sitter universe, and for $\lambda=\frac{9}{2}$, we have a matter-dominant era of the universe. \\\\
Finally, we can write the equation of state parameter i.e, $w_{eff}=-1-\frac{2\dot{H}}{3H^2}$ in the form of our dynamical variables as,
\begin{align}
    w_{eff}=-1-\frac{2Y}{3X},
\end{align}
also the decelerated parameter $q=-1-\frac{\dot{H}}{H^2}$~\cite{riess1998,Capozziello2019,Camarena2020,Rahman2023,Chaudhary2023}
 as,
\begin{align}
    q=-1-\frac{Y}{X}.
\end{align}
The value of $q$ characterizes the expansion phase of the universe: $q>0$ corresponds to decelerated expansion, $q<0$ indicates accelerated expansion and $q=0$ represents the transition between the accelerated and decelerated phase of the universe.\\\\
We will take three $f (Q,C)$ gravity models and conduct the dynamical system analysis of Bianchi-I cosmology in $f (Q,C)$. For each particular model, we first formulate the autonomous dynamical system (\ref{e:ads1}--\ref{e:ads6}). We will conduct stability analysis of each critical point (C.P), we will discuss the existence conditions of each critical point, calculate the standard density parameters $\Omega_{m}$ and $\Omega_{DE}$, decelerated parameter $q$, Equation of state (EoS) parameter $w_{eff}$, evolution equation and universe phase corresponding to each fixed point and show the phase portraits and region plot of the models. Since, in this article, we are not doing data analysis for the validation of each model. Therefore, we have not explicitly expressed the value of $H$ in Tables.

\subsection{Model: $f(Q,C) = \alpha Q^b+\beta C^{n}$}
\label{subsec:A}
\begin{figure}[h!]
    \centering
    % First subfigure
    \begin{subfigure}[b]{0.45\textwidth}
        \centering
        \includegraphics[width=\textwidth]{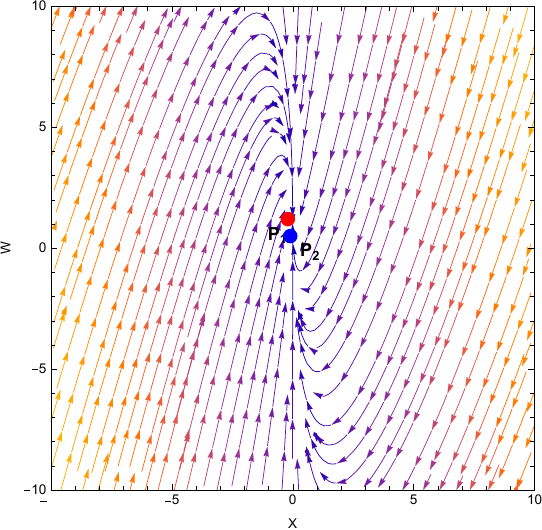}
        \caption{}
        \label{fig:sub1}
    \end{subfigure}
    % Second subfigure
    \begin{subfigure}[b]{0.45\textwidth}
        \centering
        \includegraphics[width=\textwidth]{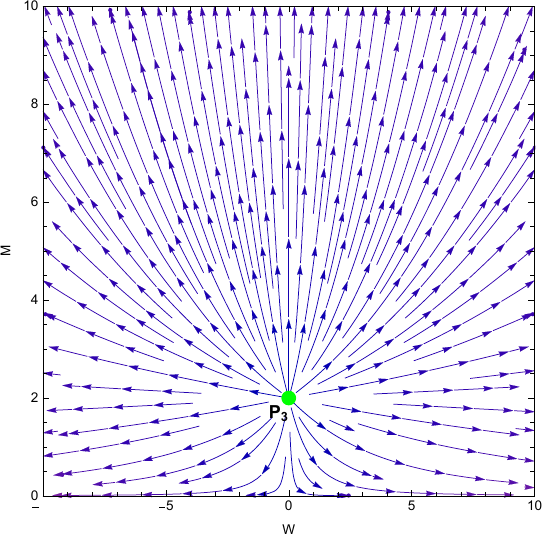}
        \caption{}
        \label{fig:sub2}
    \end{subfigure}
    
    \caption{2D phase portraits for the dynamical system for $\lambda=18,~n=-1,~b=-2.4$ (\textbf{Model} \ref{subsec:A}).}
    \label{fig1:combined}
\end{figure}

\begin{figure}[h!]
    \centering
    \includegraphics[width=0.4\textwidth]{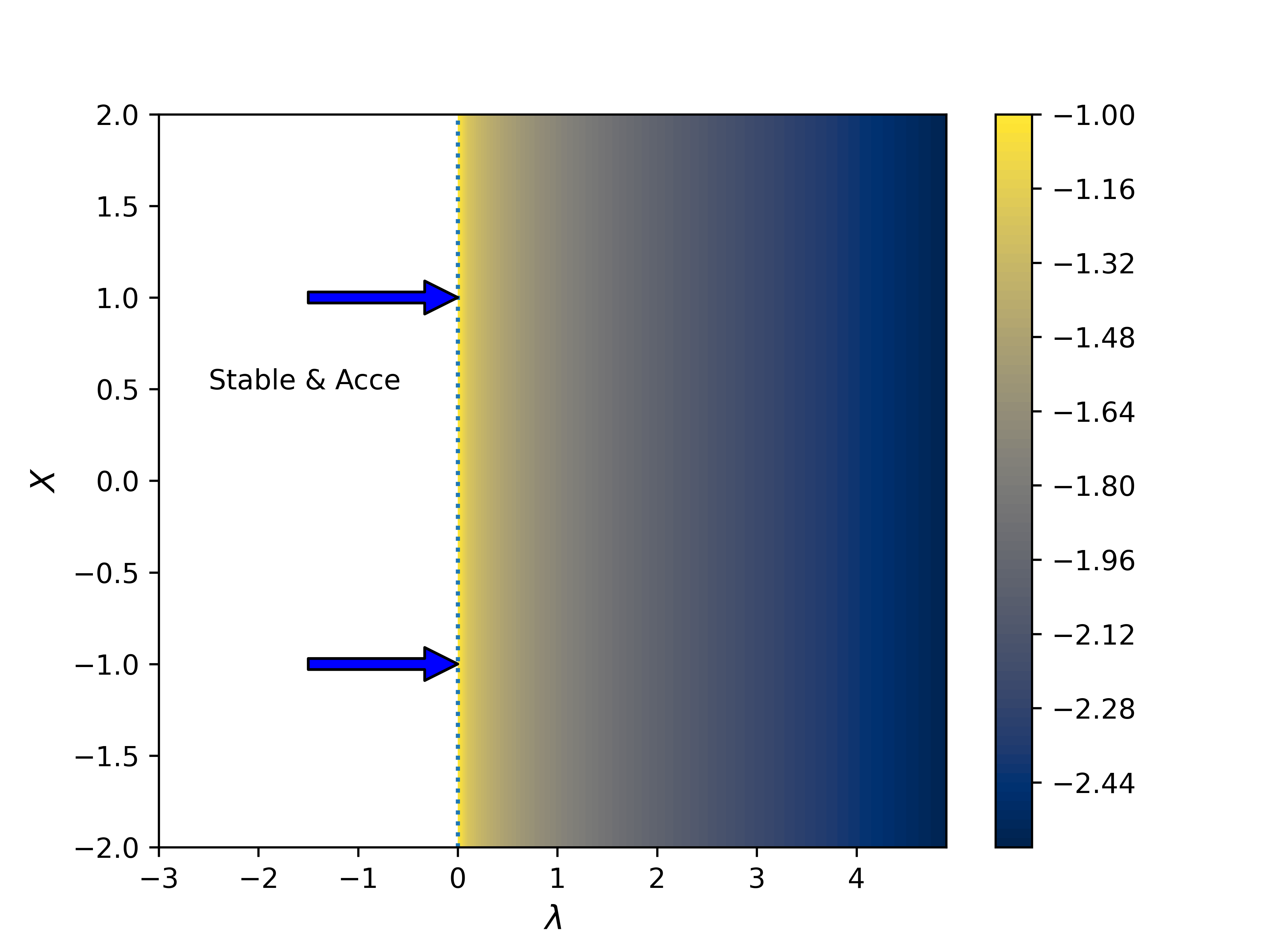}
    \caption{Region plot defining the stability and acceleration of the universe for the parameter range $\lambda$ and $X$ for C.P $P_{1}$(\textbf{Model} \ref{subsec:A})
   }
    \label{fig4}
\end{figure}

This model is thoroughly investigated in \cite{Shabani2024a}, and restores the universe evolution phases from early time to late time under certain parametric constraints. This specific form has effectively addressed the issue of late-time cosmic phenomena \cite{escamilla2020}. When $\beta=0$ and $b=1$, one can restore the standard GR equation for the matter-dominated universe. Firstly, we can express this model in terms of the dynamical variable $Z=\frac{(n-1)X(6Y+\lambda X)}{3X+Y}$ as a dependent variable. Also, we have the following expressions, $f_{Q}=\frac{2b}{2-M}\left(W+\frac{3X+Y}{n}\right)$ and $f'_{Q} = \frac{4b(b-1)(2Y + 3MX)}{X(M-2)(2bM - 2 - M)} \left(W + \frac{3X + Y}{n}\right)$. Now under this setting, our autonomous system of equations is,
\begin{align}
  X' &= \frac{(n-1)X(6Y+\lambda X)}{3X+Y}  ,
  \label{eq:X_prime} \\
  Y' &= X(\lambda-\frac{2Y^{2}}{X^{2}})+\frac{(n-1)Y(6Y+\lambda X)}{3X+Y} ,\label{eq:Y_prime} \\
%\begin{align}
 %V'=-\frac{2V(2X+Y)}{X}
%\end{align}
 W' &=-\frac{2WY}{X}-\lambda X+\frac{4bY}{X(2-M)}\left(W+\frac{3X+Y}{n}\right)-6Y+Mf'_{Q}+\frac{6bM}{2-M}\left(W+\frac{3X+Y}{n}\right) 
 \label{eq:W_prime}, \\
 M' &= \frac{nM(M-2)}{b(nW+3X+Y)}f'_{Q}-6M-\frac{2YM}{X} .
 \label{eq:M_prime}
\end{align}
The critical points and their existence are given in Table \ref{table1}.
 %and Table \ref{table4}%.
 The detailed analysis of each C.P is given below:\\\\
$P_{1}$: The critical point $P_{1}$ is a stable critical point for  $\left(\lambda>0 \land \frac{1}{2}-\frac{3}{\sqrt{2\lambda}}<b<1 \right) $  and represents the dark-energy (DE) dominance era of the universe. The eigenvalues corresponding to this fixed point, the decelerated parameter and EoS are mentioned in Table \ref{table2}. Similarly the evolution equation and universe phase are given in Table \ref{table3}. For a better understanding of the parametric range for which this fixed point is stable and represents the acceleration of the universe by DE, we have plotted the region plot given in Fig \ref{fig4} and can see that for $\lambda>0$, it gives the accelerated and stable region. The phase portrait in Fig \ref{fig:sub1}, also represents its attractor behavior.\\\\
$P_{2}$: The critical point for $\lambda=0$ represents the de Sitter solution with $q=-1$ and $\omega_{eff}=-1$. This C.P represents the standard DE-dominated era with $\Omega_{DE}=1$. The evolution and universe phase expressions are given in Table \ref{table3}. This C.P is non-hyperbolic, which means that linear stability theory (LST) is unable to describe its stability, and also the non-linear part of the above equations never vanished at zero. Hence, centre manifold theory (CMT) condition is also violated. So we use a qualitative approach, i.e. plotting the phase portrait to get stability analysis for this C.P. The phase portrait trajectories converge toward this point, as shown in Fig \ref{fig:sub1}, indicating its stable behavior.\\\\
 $P_{3}$: The $P_{3}$ is an anisotropic critical point. The EoS $\omega_{eff}=1$ implies that it is a stiff fluid. The energy density of the stiff fluid is $\rho \propto a^{-6}$, implying that it dilutes faster than other components, such as radiation and matter. Therefore, it could dominate in the very early universe, it cannot persist at late times. The decelerated parameter $q=2$ gives the decelerated expansion of the universe. The eigenvalues and values of $\Omega_{m}$ and $\Omega_{DE}$ are provided in Table \ref{table2} and Table \ref{table3}. The stability analysis gives its unstable nature, which can also be visualized in the phase portrait shown in Fig \ref{fig:sub2}. \\\\
DSA for this model has been studied for $f(T,B)$ gravity in \cite{franco2020}. So it is interesting to compare their results with our description obtained from the above analysis. It was demonstrated that setting $\lambda=0$ recovers the $\Lambda$CDM model by yielding a de Sitter solution characterized by a constant Hubble parameter. Our analysis shows that, for our isotropic critical points, in the limit where $\lambda$ vanishes, the dynamics naturally reduce to a de Sitter solution with a constant Hubble parameter, consistent with their findings.

\begin{table}[h!]
\centering
\begin{tabular}{|c|c|c|c|c|c|}
\hline
\textbf{Name of critical points}  & $X$  & $Y$  & $W$  & $M$ & \textbf{Exist for }\\ \hline
$P_{1}$ & $X_{1}$ & $\sqrt{\frac{\lambda}{2}}X_{1}$  & $-3X_{1}-\sqrt{\frac{\lambda}{2}}X_{1}$& $0$ & $X_{1}\neq 0$, $n=1$ \\ \hline
$P_{2}$ & $X_{2}$ & $0$  & $W_{2}$ &0 &$X_{2}\neq 0$, $n,W_{2},b= arbitrary, \lambda=0$ \\ \hline
$P_{3}$& $X_{3}$ & $-3X_{3}$ & $0$& $M_{3}$ & $X_{3}\neq 0$, $n=1$, $M_{3}\neq 0$ ,$\lambda=\frac{2Y_{3}^2}{X_{3}^{2}}$ \\ \hline
\end{tabular}
\caption{}
\label{table1}
\end{table}
\begin{table}[h!]
    \centering
    \resizebox{\textwidth}{!}{%
        \begin{tabular}{|c|c|c|c|c|}
            \hline
            \textbf{C.P} &\textbf{Eigenvalues} &\textbf{ Stability} & $q$ & $w_{eff}$\\
            \hline
            $P_{1}$ & $(0,-\sqrt{2\lambda}+b\sqrt{2\lambda},-6+2\sqrt{2\lambda}-2b\sqrt{2\lambda},-2\sqrt{2\lambda})$ & stable for $\left(\lambda>0 \land \frac{1}{2}-\frac{3}{\sqrt{2\lambda}}<b<1 \right) $  & $-1-\sqrt{\frac{\lambda}{2}}$ & $-1-\frac{\sqrt{2\lambda}}{3}$ \\
            \hline
            $P_{2}$ & $(0,0,0,-6)$ & non-hyperbolic  & $-1$ & $-1$ \\
            \hline
            $P_{3}$ & $(0,\frac{6(b-1)(M_{3}+2)}{2bM_{3}-2-M_{3}},12,\frac{12(b-1)(4-4M_{3}-M_{3}^2 +2bM_{3}^2)}{(2bM_{3}-2-M_{3})^2})$ & unstable & 2 & 1 \\
            \hline
        \end{tabular}%
    }
    \caption{}
    \label{table2}
\end{table}
\begin{table}[h!]
    \centering
    \begin{tabular}{|c|c|c|c|c|}
        \hline
        \textbf{C.P}& $\Omega_{m}$& $\Omega_{DE}$& \textbf{Evolution eqs} & \textbf{Universe phase} \\ \hline
        $P_{1}$  &0&  1& $\dot{H}=\sqrt{\frac{\lambda}{2}} H^{2}$     & $a \sim t^{-\sqrt{\frac{2}{\lambda}}}$        \\ \hline
       $P_{2}$   &0&1   & $\dot{H}=0$     &$a \sim e^{ct}$       \\ \hline
       $P_{3}$  & $2b$& $1-2b $  & $\dot{H}=-3H^{2}$      & $a \sim t^{\frac{1}{3}}$     \\ \hline
    \end{tabular}
    \caption{}
    \label{table3}
\end{table}

\subsection{Model: $f(Q,C) = f_{0} Q^b C^{n}$}
\label{subsec:B}
\begin{figure}[h!]
    \centering
    % First subfigure
    \begin{subfigure}[b]{0.45\textwidth}
        \centering
        \includegraphics[width=\textwidth]{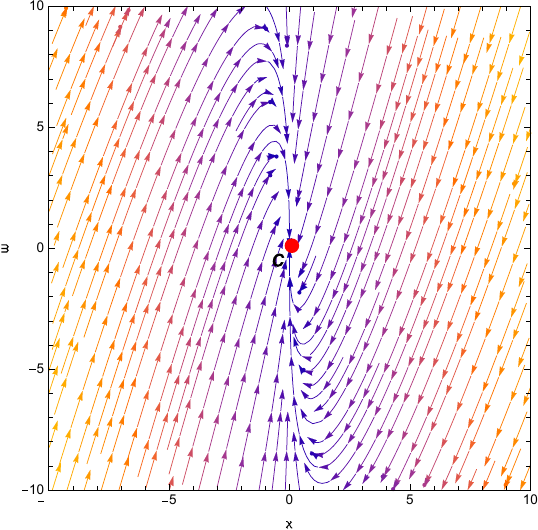}
        \caption{}
        \label{fig:sub3}
    \end{subfigure}
    % Second subfigure
    \begin{subfigure}[b]{0.45\textwidth}
        \centering
        \includegraphics[width=\textwidth]{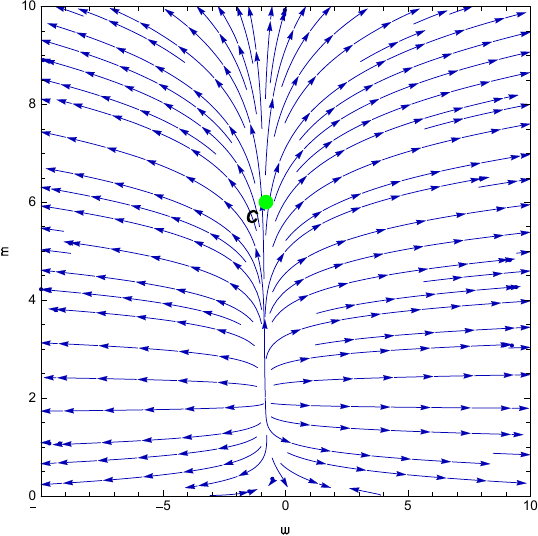}
        \caption{}
        \label{fig:sub4}
    \end{subfigure}
    
    \caption{Phase portraits in 2D for the dynamical system at $\lambda=14,~n=-1,~b=0.21$ (\textbf{Model} \ref{subsec:B}).}
    \label{fig2:combined}
\end{figure}

We have considered the product form of $f(Q,C)$ to obtain the characteristics of power-law behaviors observed in various cosmological epochs throughout the evolution history of the universe. Here $f_{0}$, $b$ and $n$ are constants. In Ref. \cite {Lohakare2024}, a cosmological data analysis was performed for this model, indicating a universe dominated by dark energy and undergoing late-time accelerated expansion. Stability analysis was conducted in \cite{franco2020}, showing that the $\Lambda$CDM model is recovered in the GR limit. We get the following form of the variable $z=\frac{2b(2y f_{Q}+mxf'_{Q}+3mxf_{Q})}{(2-m)f_{Q}}+\frac{(n-1)x(6y+\lambda x)}{3x+y}$ for this particular model and treated as a dependent variable. For this model, we have  $f_{Q}=\frac{2bw}{2-m}$ and $f'_{Q}=\frac{4b(b-1)w}{(m-2)(2bm-m-2)}(\frac{2y}{x}+3m)-\frac{2bnw}{(2bm-2-m)}(\frac{\lambda x+6y}{3x+y})$, so the autonomous dynamical system (\ref{e:ads1}--\ref{e:ads6}) can be written as,
%For this $z=\frac{(n-1)x(6y+\lambda x)}{3x+y}$,~ $f_{Q}=\frac{2bw}{2-m}$,~ $f'_{Q}=\frac{4b(b-1)w}{(m-2)(2bm-m-2))}(\frac{2y}{x}+3m)$
\begin{align}
  x' &= \frac{2b(2y f_{Q}+mxf'_{Q}+3mxf_{Q})}{(2-m)f_{Q}}+\frac{(n-1)x(6y+\lambda x)}{3x+y},
  \label{eq:x_prime} \\
  y' &= x\left(\lambda-\frac{2y^{2}}{x^{2}}\right)+\frac{4by^2}{x(2-m)}+\frac{ymf'_{Q}}{w}+\frac{6bym}{2-m}+\frac{(n-1)y(6y+\lambda x)}{3x+y},
  \label{eq:y_prime} \\
%\begin{align}
 %V'=-4V-\frac{2YV}{X}
%\end{align}
 w' &= -\frac{2wy}{x} - \lambda x+\frac{4byw}{x(2-m)}-6y+mf'_{Q}+\frac{6bmw}{2-m} ,
 \label{eq:w_prime} \\
 m' &= -\frac{mf'_{Q}(2-m)}{bw}-6m-\frac{2ym}{x}.
 \label{eq:m_prime}
\end{align}
We have calculated the critical points and their existence listed in Table \ref{table7}. %and Table \ref{table10}.
Let us discuss the details of each critical point.\\\\
$C_{1}$: The critical point represents the de Sitter solution with $\omega_{eff}=-1$ and $q=-1$. It indicates the standard dark energy-dominated era at $x_{1}~ and~ w_{1}=0$. The evolution and universe phase expressions are given in Table \ref{table9}. This C.P is non-hyperbolic which implies the failure of the stability theory. Moreover, the system of equations does not fulfill the CMT criteria of vanishing the non-linear part of the system of equations after separation. However, the phase space analysis in Fig \ref{fig:sub3}, provides a stable nature of this fixed point.\\\\
$C_{2}$: This critical point is an anisotropic critical point. It gives the stiff fluid with EoS $\omega_{eff}=-1$. The universe phase and evolution equation corresponding to this are mentioned in Table \ref{table9}. The linear stability analysis provides its saddle behavior that can also be verified by the phase portrait in Fig \ref{fig:sub4}, indicating the saddle, and thus the unstable nature of this C.P.\\\\
A DSA for this example was conducted within the framework of $f(Q,B)$ gravity in \cite{Lohakare2024}. Their study identified a unique stable critical point that corresponds to a de Sitter phase of the universe, a result that aligns with our isotropic critical point.

\begin{table}[h!]
\centering
\begin{tabular}{|c|c|c|c|c|c|}
\hline
\textbf{Name of critical points}  & $x$  & $y$ & $w$  & $m$ & \textbf{Exist for} \\ \hline
$C_{1}$ & $x_{1}$ & $0$   & $w_{1}$& $0$ & $x_{1}\neq 0$, $n,b,w_{1}=arbitrary$, $\lambda=0$ \\ \hline
$C_{2}$ & $x_{2}$ & $-3x_{2}$   & $\frac{3b}{b-1}$& $\frac{-2y_{2}}{x_{2}}$ & $x_{2}\neq 0$, $n=1$, $\lambda=18(1-b), b=arbitrary$\\ \hline
\end{tabular}
\caption{}
    \label{table7}
\end{table}

\begin{table}[h!]
\centering
%\scriptsize
\begin{tabular}{|c|c|c|c|c|}
\hline
\textbf{C.P} & \textbf{Eigenvalues} & \textbf{Stability} & $q$ & $w_{eff}$ \\
\hline
$C_{1}$ & $(0,0,0,-6 )$ & non-hyperbolic  & $-1$ & $-1$ \\
\hline
$C_{2}$ & $\left(-\frac{24b}{3(2b-1)-1},-\frac{216b(b-1)}{3(2b-1)-1},\frac{18b(3(4b-1)-1)}{3(2b-1)-1},\frac{144(b-1)}{3(2b-1)-1}\right)$ & saddle & $2$ & $1$ \\
\hline
\end{tabular}
\caption{}
    \label{table8}
\end{table}

\begin{table}[h!]
    \centering
    \begin{tabular}{|c|c|c|c|c|}
        \hline
        \textbf{C.P} & $\Omega_{m}$& $\Omega_{DE}$& \textbf{Evolution eqs} & \textbf{Universe phase} \\ \hline
        $C_{1}$    & $w_{1}(2b-1)-3x_{1}$ & $1-w_{1}(2b-1)+3x_{1}$&$\dot{H}=0$     & $a \sim e^{ct}$        \\ \hline
       $C_{2}$      &$6bx_{2}+\frac{6b^2}{b-1}-\frac{3b}{b-1}$ &$1-6bx_{2}-\frac{6b^2}{b-1}+\frac{3b}{b-1}$ & $\dot{H}=-3 H^{2}$     &$a \sim t^{\frac{1}{3}}$       \\ \hline
    \end{tabular}
    \caption{}
    \label{table9}
\end{table}

\subsection{Model: $f(Q,C) = \zeta Q+\alpha C log(C)$}
\label{subsec:C}
\begin{figure}[h!]
    \centering
    % First subfigure
    \begin{subfigure}[b]{0.45\textwidth}
        \centering
        \includegraphics[width=\textwidth]{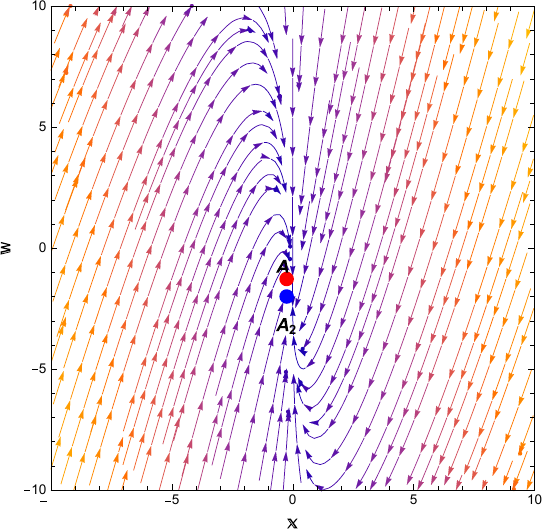}
        \caption{}
        \label{fig:sub5}
    \end{subfigure}
    % Second subfigure
    \begin{subfigure}[b]{0.45\textwidth}
        \centering
        \includegraphics[width=\textwidth]{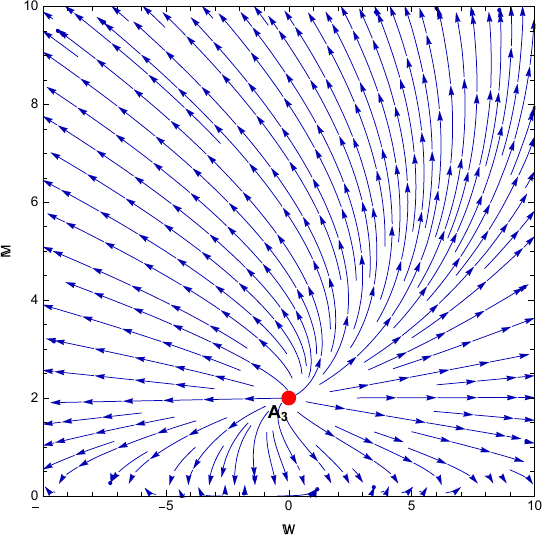}
        \caption{}
        \label{fig:sub6}
    \end{subfigure}
    
    \caption{2D phase portraits for the dynamical system for $\lambda=18,~\alpha=-2,~\zeta=-2.4$  (\textbf{Model} \ref{subsec:C}).}
    \label{fig1:combined}
\end{figure}

\begin{figure}[ht]
    \centering
    \includegraphics[width=0.4\textwidth]{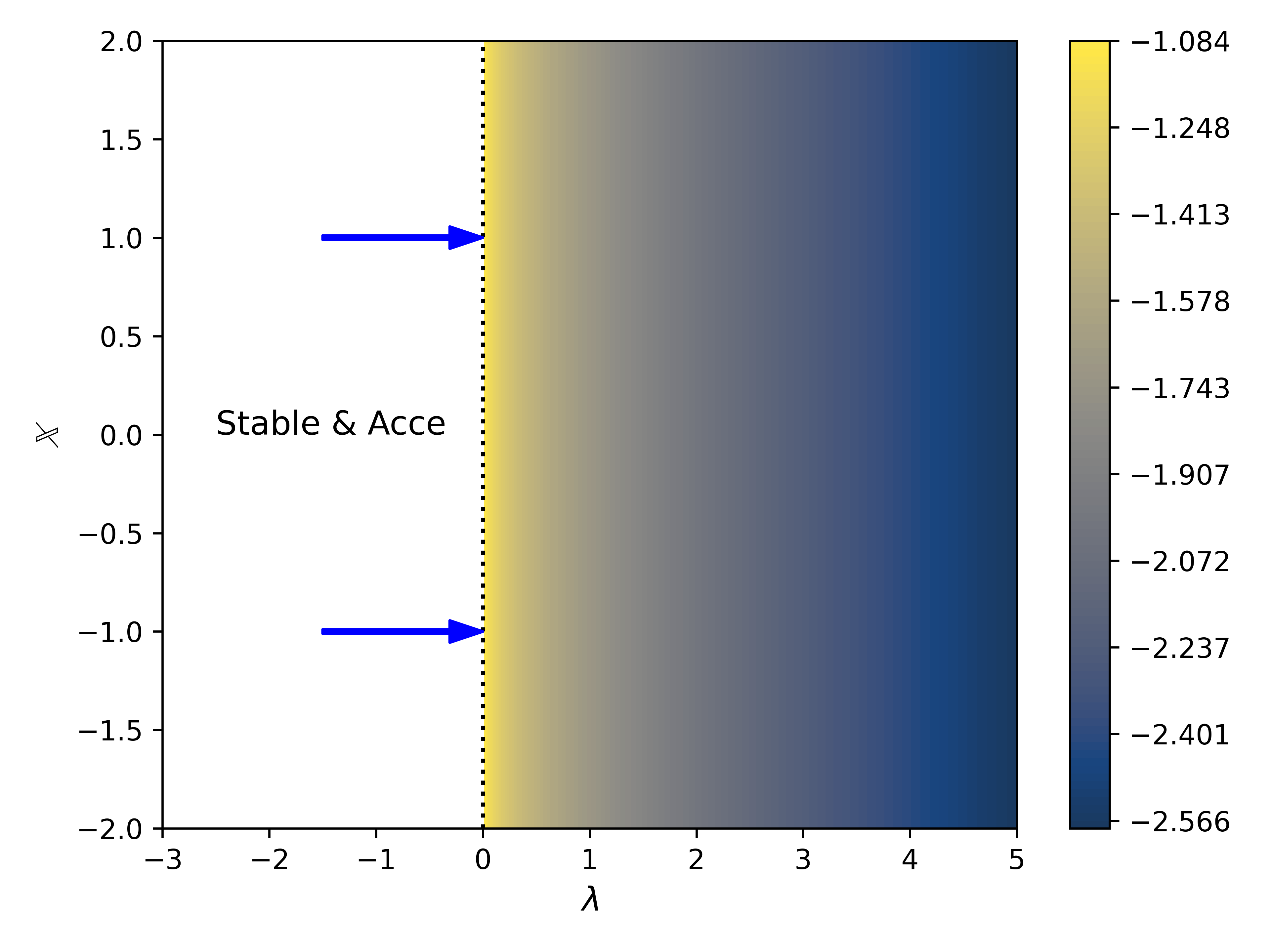}
    \caption{Region plot defining the stability and acceleration of the universe for the parameter range $\lambda$ and $\mathbb{X}$ for C.P $A_{1}$(\textbf{Model} \ref{subsec:C})
   }
    \label{fig14}
\end{figure}

This model has proven effective in resolving the late-time cosmic phenomena \cite{escamilla2020,bamba2011}. By exploring the behavior of the critical points, particularly in the absence of a scalar field, one may get a deeper understanding of cosmological evolution through distinct phases \cite{Paliathanasis2021a}. The dynamical variable $Z$ can be written in the following way $\mathbb{Z}= \alpha(\frac{ 6\mathbb{Y}+\lambda \mathbb{X}}{3\mathbb{X}+\mathbb{Y}})$. For this model, $f_{Q}=\zeta$ and $f'_{Q}=0$. So our autonomous dynamical system for this model is provided as,
\begin{align}
  \mathbb{X}' &= \alpha(\frac{ 6\mathbb{Y}+\lambda \mathbb{X}}{3\mathbb{X}+\mathbb{Y}}),
  \label{eq:xx_prime} \\
  \mathbb{Y}' &= \mathbb{X}(\lambda-\frac{2\mathbb{Y}^{2}}{\mathbb{X}^{2}})+\frac{\alpha \mathbb{Y}(6\mathbb{Y}+\lambda \mathbb{X})}{\mathbb{X}(3\mathbb{X}+\mathbb{Y})},
  \label{eq:yy_prime} \\
%\begin{align}
 %V'=-\frac{2V(2X+Y)}{X}
%\end{align}
 \mathbb{W}' &= -\frac{2\mathbb{W}\mathbb{Y}}{\mathbb{X}}-\lambda \mathbb{X}+\frac{2\mathbb{Y}\zeta}{\mathbb{X}}-6\mathbb{Y}+3\mathbb{M}\zeta ,
 \label{eq:ww_prime} \\
 \mathbb{M}' &= -6\mathbb{M}-\frac{2\mathbb{Y}\mathbb{M}}{\mathbb{X}}. \label{eq:mm_prime}
\end{align}
The obtained critical points are mentioned in Table \ref{table13}. 
% and Table \ref{table16}.
We shall explain each C.P in detail as follows:\\\\
$A_{1}$: The critical point $A_{1}$ indicates the DE dominance epoch of the universe at $\zeta=0$. It is a stable point with stability condition $\mathbb{X}_{1} \in \mathbb{R} \land \lambda>0$. The eigenvalues, decelerated parameter, EoS, evolution equation, and universe phase for this C.P with parameter $\lambda$ dependence are written in Table \ref{table14} and Table \ref{table15} respectively. The region plot in Fig \ref{fig14}, effectively captures the range of parameters where this fixed point is stable and accounts for the universe's acceleration driven by DE. The trajectories are converging toward this point, validating its stable nature, as shown in Fig \ref{fig:sub5}.\\\\
$A_{2}$: For $\lambda=0$, we have the critical point $A_{2}$, which corresponds to the de Sitter solution with $\omega_{eff}=-1$ and $q=-1$. It provides the standard DE dominated era at $\zeta,\mathbb{W}_{2}, and~ \mathbb{X}_{2} =0$. 
%The evolution and universe phase are given in the Table \ref{table15}. 
Similar to the first model, this point is non-hyperbolic as linear stability does not work. Additionally, the system of equations does not satisfy the CMT condition. But the phase space analysis in Fig \ref{fig:sub5}, represents the stable nature of this fixed point..\\\\
$A_{3}$: It represents an anisotropic critical point. It describes as a stiff fluid with $\omega_{eff}=1$. The decelerated parameter represents the decelerated expansion of the universe with $q=2$. 
%In Table \ref{table15}, we have the universe phase and evolution expression for this fixed point.
This C.P is unstable. The phase portrait analysis is given in Fig \ref{fig:sub6}, which demonstrates its unstable nature.\\\\
A DSA for this particular model has also been investigated under the framework of $f(T,B)$ gravity in \cite{Kadam2023b}. Their analysis revealed that a de Sitter solution emerges for vanishing $\lambda$, a result that aligns with our analysis of the isotropic critical points.

\begin{table}[h!]
\centering
\begin{tabular}{|c|c|c|c|c|c|}
\hline
\textbf{Name of critical points}  & $\mathbb{X}$  & $\mathbb{Y}$  & $\mathbb{W}$  & $\mathbb{M}$ & \textbf{Exist for} \\ \hline
$A_{1}$ & $\mathbb{X}_{1}$ & $\sqrt{\frac{\lambda}{2}}$$\mathbb{X}_{1}$  & $-\sqrt{\frac{\lambda}{2}}$ $\mathbb{X}_{1}$+$\zeta$-3$\mathbb{X}_{1}$& $0$ & $\mathbb{X}_{1}\neq 0$, $\alpha=0$ \\ \hline
$A_{2}$ & $\mathbb{X}_{2}$ & $0$ & $\mathbb{W}_{2}$ & $0$ & $\mathbb{X}_{2} \neq 0, \mathbb{W}_{2}, \alpha, \zeta=arbitrary, \lambda= 0$ \\ \hline

$A_{3}$& $\mathbb{X}_{3}$ & $-3\mathbb{X}_{3}$  & $ \zeta(1-\frac{\mathbb{M}_{3}}{2})$& $\mathbb{M}_{3}$ & $\mathbb{X}_{3}\neq 0$, $\alpha=0$, $\mathbb{M}_{3}\neq 0$ ,$\lambda=\frac{2\mathbb{Y}_{3}^2}{\mathbb{X}_{3}^{2}}$ \\ \hline
\end{tabular}
\caption{}
    \label{table13}
\end{table}

\begin{table}[h!]
    \centering
\begin{tabular}{|c|c|c|c|c|}
\hline
\textbf{C.P} & \textbf{Eigenvalues} & \textbf{Stability}  & $q$& $w_{eff}$\\
\hline
$A_{1}$ & $(0,-6-\sqrt{2\lambda},-2\sqrt{2\lambda},-\sqrt{2\lambda})$ & stable for $\mathbb{X}_{1} \in \mathbb{R} \land \lambda>0$& $-1 - \sqrt{\frac{\lambda}{2}}$ & $-1 - \frac{\sqrt{2\lambda}}{3}$ \\
\hline
$A_{2}$ & $(0,0,0,-6)$ & non-hyperbolic &  $-1 $ & $-1$ \\
\hline
$A_{3}$ & $(0,0,0,6)$&  unstable & 2 & 1 \\
\hline
\end{tabular}
\caption{}
    \label{table14}
\end{table}

\begin{table}[h!]
    \centering
    \begin{tabular}{|c|c|c|c|c|}
        \hline
        \textbf{C.P} & $\Omega_{m}$& $\Omega_{DE}$& \textbf{Evolution eqs} & \textbf{Universe phase} \\ \hline
        $A_{1}$    & $\zeta$& $1 - \zeta$ &$\dot{H}=\sqrt{\frac{\lambda}{2}} H^{2}$     & $a \sim t^{-\sqrt{\frac{2}{\lambda}}}$        \\ \hline
       $A_{2}$      & $2\zeta-\mathbb{W}_{2}-3\mathbb{X}_{2}$& $1-2\zeta+\mathbb{W}_{2}+3\mathbb{X}_{2}$ &$\dot{H}=0$     &$a \sim e^{ct}$       \\ \hline
       $A_{3}$      &  $\mathbb{W}_{3}$ &$1 - \mathbb{W}_{3}$ &$\dot{H}=-3H^{2}$      & $a \sim t^{\frac{1}{3}}$     \\ \hline
    \end{tabular}
    \caption{}
    \label{table15}
\end{table}

%%%%%%%%%%%%%%%%%%%%%%%%%%%%%%%%%%%%%%%%%%%%%%%%%%%%%%%%%%%%%%%%%%%%%%%%%%%%
\section{Conclusion}\label{sec5}
In this paper, we have considered the Bianchi-I cosmology in $f(Q,C)$ theory and have conducted a dynamical system analysis. In this study, the universe is considered to be filled with matter fluid.  Firstly, we have formulated the general dynamical system (\ref{e:ads1}--\ref{e:ads6}), independent of any specific choice of the function $f(Q,C)$. Then, we considered three different models of $f(Q,C)$ theory to analyze their dynamical systems. In each model, we have computed the critical points and explained their characteristics in detail.\\\\
The first model that we have considered is $f(Q,C) = \alpha Q^b+\beta C^{n}$. The autonomous system of equations of this model is given in (\ref{eq:X_prime}--\ref{eq:M_prime}). The stability analysis of each critical point is performed based on the eigenvalues, obtained from the Jacobian matrix at each critical point. For this model, we have observed that the critical point $P_{1}$ is stable and gives the DE dominance epoch of the universe. The range for the parameter $\lambda$ in which this C.P is stable and gives the acceleration of the universe by DE is plotted in Fig \ref{fig4}. The phase portrait in Fig \ref{fig:sub1} also depicts its stable nature. The $P_{2}$ is a critical point, which we have obtained for $\lambda=0$ and it gives the de Sitter solution and dark energy dominance era with $\Omega_{DE}=1$. The phase portrait in Fig \ref{fig:sub1}, confirms its stable behavior. $P_{3}$ is an anisotropic fixed point, which represents the stiff fluid and decelerated expansion of the universe with $q=2$. The linear stability provides its unstable state as highlighted in Fig \ref{fig:sub2}. \\\\
The second model we have studied is $f(Q,C) = f_{0} Q^b C^{n}$. The equations (\ref{eq:x_prime}-\ref{eq:m_prime}) provide the autonomous dynamical system for this model. The possible viable critical points and corresponding existence conditions of every fixed point are written in Table \ref{table7}. % and Table \ref{table10}.
We have noticed that the critical point $C_{1}$ represents de Sitter universe with stable nature as shown in Fig \ref{fig:sub3}. The anisotropic critical point $C_{2}$, gives stiff fluid with $\omega_{eff}=1$. The saddle, and thus the unstable state of this point, is given in Fig \ref{fig:sub4}.\\\\
The third model we have considered is $f(Q,C) = \zeta Q+\alpha C log(C)$. The autonomous dynamical system for this particular model is given by equations (\ref{eq:xx_prime}-\ref{eq:mm_prime}).  The critical points and the conditions for their existence of this particular model are discussed in Table \ref{table13}. % and Table \ref{table16}.
 We have analyzed that, the critical point $A_{1}$ gives the DE dominance universe with a stable behavior as presented in Fig \ref{fig:sub5}. The stable and accelerated expansion of the universe due to DE, corresponding to this C.P, for the range of $\lambda$, is displayed in Fig \ref{fig14}. The critical point $A_{2}$, gives de Sitter universe with a stable state, and the anisotropic fixed point $A_{3}$ represents a stiff fluid with unstable behavior.\\\\
The analysis of the above three models with anisotropic effect gives compatible and some interesting evolution scenarios of the universe from early time to late time. By doing the cosmological observational data analysis of these critical points and flexibility in the model parameters, one can get more deeper insight.

%%%%%%%%%%%%%%%%%%%%%%%%%%%%%%%%%%%%%%%%%%%%%%%%%%%%%%%%%%%%%%%%%
%\section{Acknowledgement}
%The research has been carried out under Universiti Tunku Abdul Rahman Research Fund project \\IPSR/RMC/UTARRF/2023-C1/A09 provided by Universiti Tunku Abdul Rahman.

%%%%%%%%%%%%%%%%%%%%%%%%%%%%%%%%%%%%%%%%%%%%%%%%%%%%%%%%%%%%%%%%%
%%%%%%%%%%%%%%%%%%%%%%%%%%%%%%%%%%%%%%%%%%%%%%%%%%%%%%%%%%%%%%%%%

\end{document}